\documentstyle[9lomcon,epsfig]{article}

\begin{document}

\title{COLOR RECONNECTION AND BOSE-EINSTEIN CORRELATIONS AT LEP2}

\author{ Th. Ziegler \footnote{e-mail: Thomas.Ziegler@cern.ch}}

\address{Institut f\"ur Physik, Johannes Gutenberg Universit\"at,
Staudingerweg 7, 55099 Mainz, Germany}


\maketitle\abstracts{
Recent investigations of final state interactions of $W^+W^-$ events 
in $e^+e^-$ collisions up to center of mass energies $\sqrt{s} = 189$ $GeV$
at LEP2 are reviewed. The data were used to look for color reconnection and
Bose-Einstein correlations between the two color singlets of fully hadronic
W events.}

\section{Motivation}
In high energy $e^+e^-$ collisions at LEP the W bosons are produced in
pairs. 
As the hadronisation scale is much larger than the distance of the W bosons
at their primary decay vertices, the decay products have a significant
space-time overlap and the two systems may interfere during the
hadronisation phase. One important consequence is a possible shift in the
invariant mass of the reconstructed W bosons in the fully hadronic channel.
Two types of final state interactions are investigated and will be reviewed
seperately: color reconnection and Bose-Einstein correlations.\\
The results are based on $\approx 55$ $pb^{-1}$ at $\sqrt{s} = 183$ $GeV$ and
$\approx 173$ $pb^{-1}$ at $\sqrt{s} = 189$ $GeV$ per LEP collaboration.

\section{Color Reconnection (CRC)}
Color reconnection (CRC) leads to a rearrangement of the color flow between
the hadronic decay products of the W bosons. 
That this may cause a possibly significant shift in the W mass measurement
was first suggested by Sj\"ostrand and Khoze [\cite{th_sk1}].
In the perturbative case the color flow of the primary quarks is rearranged
and possible effects can be estimated by perturbative QCD. 
The shift of the W mass is predicted to 
$\Delta M_W \le \mathcal{O}$$(5 MeV)$
and thus negligible.
The effect due to the rearrangement of the color flow of the secondary decay
products has to be investigated by MC studies and non-perturbative QCD models
have to be used. The consequence for the W mass shift is estimated to be of
the order of $\Delta M_W \le \mathcal{O}$$(50 MeV)$. 
The following review concentrates on the latter case.

\subsection{Inclusive Mean Charged Particle Multiplicity}
\begin{figure}[t]
\begin{minipage}{5.4cm}
{\footnotesize
\begin{tabular}{|c|c|} \hline
\rule[-2mm]{0mm}{5mm} & $\Delta \langle N_{ch} \rangle = 
              \langle N_{ch}^{4q} \rangle
              - 2 \langle N_{ch}^{qq\ell \nu} \rangle$ \\
              \hline \hline
{\scriptsize ALEPH}      &  0.47 $\pm$ 0.44 $\pm$ 0.26 \\ \hline
{\scriptsize L3}         & -0.23 $\pm$ 0.40 $\pm$ 0.52 \\ \hline
{\scriptsize OPAL}       &  0.7  $\pm$ 0.8  $\pm$ 0.6  
              \footnotemark
	      \\ \hline
{\scriptsize DELPHI}     & -0.92 $\pm$ 0.66 
              \footnotemark
              \\ \hline
\rule[-2mm]{0mm}{5mm}   & $\langle N_{ch}^{4q} \rangle / 2 \cdot 
              \langle N_{ch}^{qq\ell \nu} \rangle$ \\ \hline
{\scriptsize DELPHI}     & 0.977 $\pm$ 0.017 $\pm$ 0.027 \\ \hline
\end{tabular}
}
\caption[]{\small \\
a) Studies of the mean charged particle multiplicity of the
LEP collaborations at $\sqrt{s} = 189~GeV$ (table above).\\
b) Inclusive charged particle momentum distributions and comparison with MC
predictions with and without CRC models (figure on the right)}
\label{fig_crc_aleph1}
\label{tab_nchtab}
\end{minipage}
\begin{minipage}{5cm}
\epsfig{file=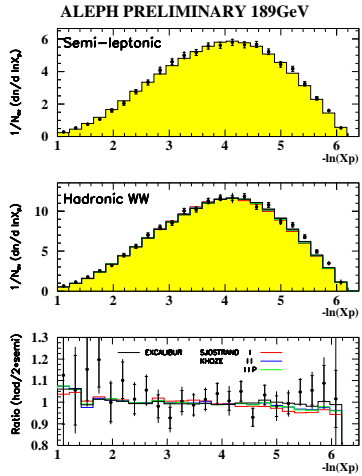,width=5cm,clip=}
\end{minipage}
\end{figure}

To investigate effects of CRC simple observables
like the inclusive charged multiplicity $N_{ch}$ are obvious candidates.
There have been models [\cite{th_eg1}] which predict a $\approx 10 \%$
effect of CRC on the mean charged multiplicity
$\langle N_{ch}^{4q} \rangle$ in the $W^+W^- \rightarrow q\bar{q}q\bar{q}$
channel which were excluded in earlier analysis [\cite{crc_opal1}, a.o.]  but
encouraged the interest in these kind of studies.\\
As a reference for $\langle N_{ch}^{4q} \rangle$ the mean charged multiplicity
$\langle N_{ch}^{qq\ell \nu} \rangle$ of the hadronic part of the
semileptonic channel is used.
Typically the difference 
$\Delta \langle N_{ch} \rangle = \langle N_{ch}^{4q} \rangle - 
 2 \cdot \langle N_{ch}^{qq\ell v} \rangle$
is chosen as an observable and should be 0 for no observed effect.
The results of the four collaborations
[\cite{be_aleph1} - \cite{be_opal1}] in figure \ref{fig_crc_aleph1}a
should not be compared directly with each other due to different unfolding
and correction procedures.
MC studies with physically reasonable CRC models predict a shift in the mean
charged multiplicity of the fully hadronic channel of 
$| \Delta \langle N_{ch} \rangle ^{MC} | \le 0.3$. 
Therefore all the results are still compatible with standard model
predictions.
% set footnotecounter 1 back
\addtocounter{footnote}{-1}
\footnotetext{result for $\sqrt{s} = 183~GeV$}
\stepcounter{footnote}
\footnotetext{extracted from the numbers given in the DELPHI paper}

\subsection{Inclusive Charged Particle Momentum Distributions}
\begin{figure}[t]
\begin{minipage}[t]{5.1cm}
\epsfig{file=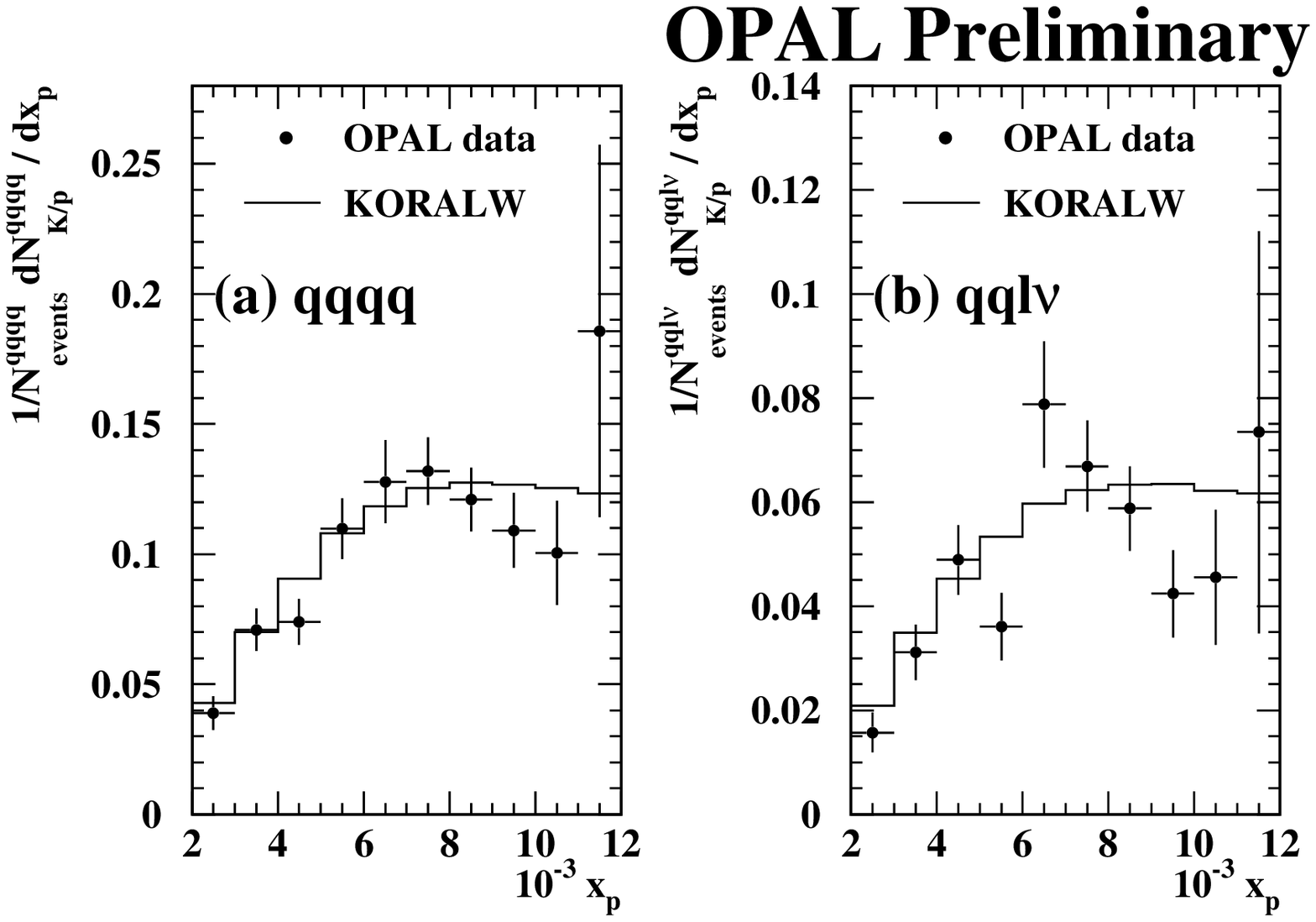,width=5.1cm,clip=}
\end{minipage}
\begin{minipage}[b]{5.3cm}
\epsfig{file=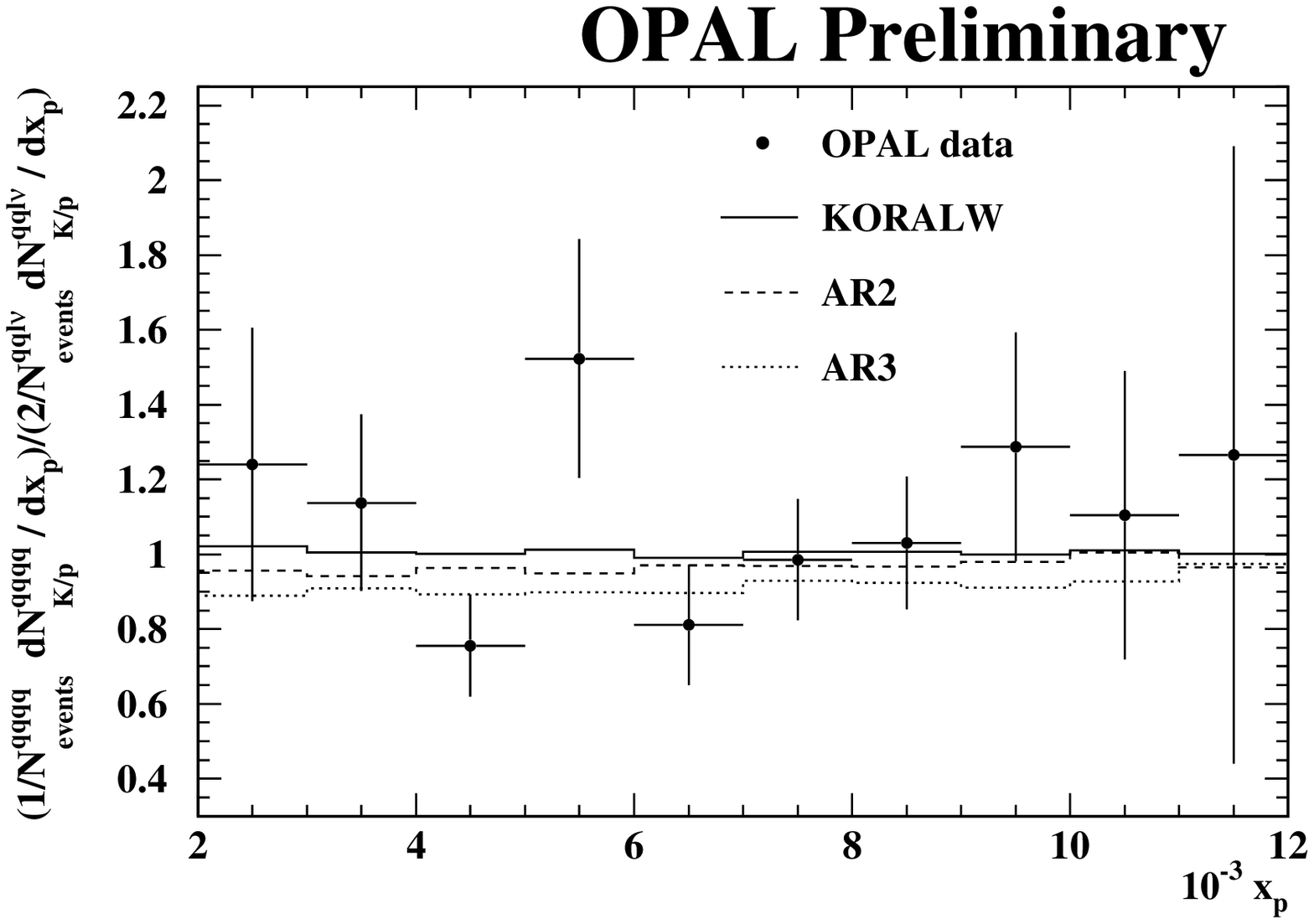,width=5.3cm,clip=}
\end{minipage}
\vspace{-0.6cm}
\caption{\small The charged low momentum distribution of charged Kaons and
Protons and their ratio compared to MC predictions with and without CRC
models.} 
\label{crc_opal1}
\end{figure}

From MC studies an effect of CRC [\cite{th_sk2}], especially in the low
momentum region of the hadronic WW events, is expected.
ALEPH [\cite{crc_aleph1}], DELPHI [\cite{crc_delphi1}] and OPAL
[\cite{crc_opal1}] performed similar analysis on momentum distributions.
However no evidence for CRC within the statistical significance could be
found. As an example the $\xi$ distribution of ALEPH is shown in figure
\ref{fig_crc_aleph1}b where $\xi$ is the transformed normalized momentum of
the charged particles:
$\xi = - ln(x_p) = - ln(\frac{p}{\sqrt{s}/2})$. 
A comparison of the hadronic and the semileptonic channel shows no evidence
for color reconnection. Even more low momentum particles are observed whereas
less are expected by the 3 shown CRC models of Sj\"ostrand and Khoze.
Further MC studies claimed [\cite{th_sk2}], that the effect should be
enlarged for low momentum heavy particles in the context of Sj\"ostrands CRC
models. 
DELPHI [\cite{crc_delphi1}] and OPAL [\cite{crc_opal2}] performed analyses
on the low momentum distributions of charged Kaons and Protons. The result
of OPAL is shown in figure \ref{crc_opal1}.
In both channels some discrepancies between the data and the KORALW standard
MC are found.
Nevertheless the ratio
$R = N_{K,p}^{4q} / (2 \cdot N_{K,p}^{qq\ell \nu})$ prefers the MC without
CRC. 
For comparison 2 ARIADNE CRC models are shown.
The ratio of the production rates in the momentum region
$0.002 \le x_p \le 0.012$ was found to be\\
\centerline{$R(183~GeV) = 0.91 \pm 0.13 \pm 0.08$}
\centerline{$R(189~GeV) = 1.11 \pm 0.08 \pm 0.06$}
For both energies the result is consistent with the expectation of standard
QCD models without CRC.

\subsection{Particle and Energy Flow Distributions between Jets}
\begin{figure}[t]
\begin{minipage}{5cm}
\epsfig{file=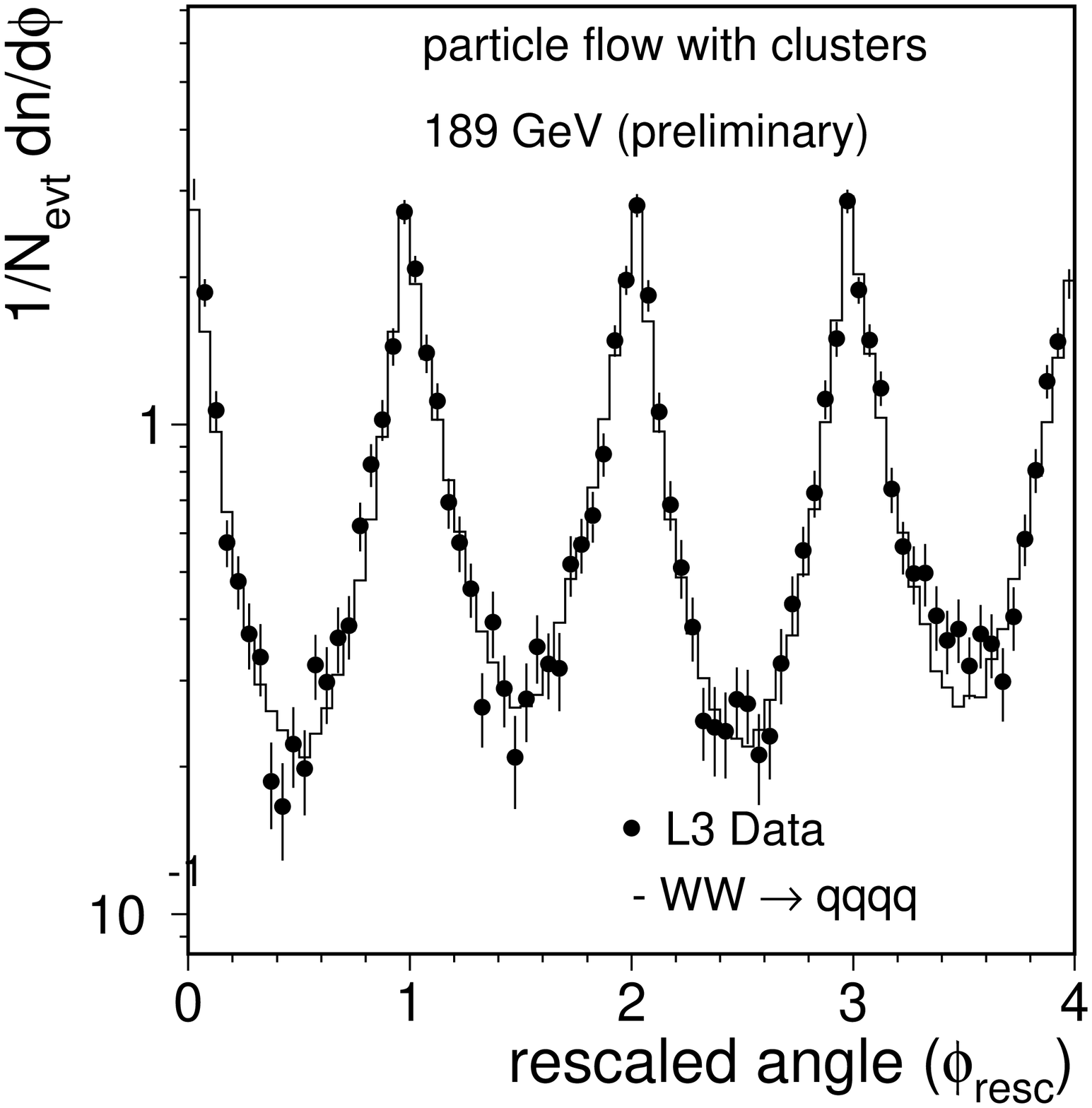,width=5.2cm,clip=}
\end{minipage}
\begin{minipage}{5cm}
\epsfig{file=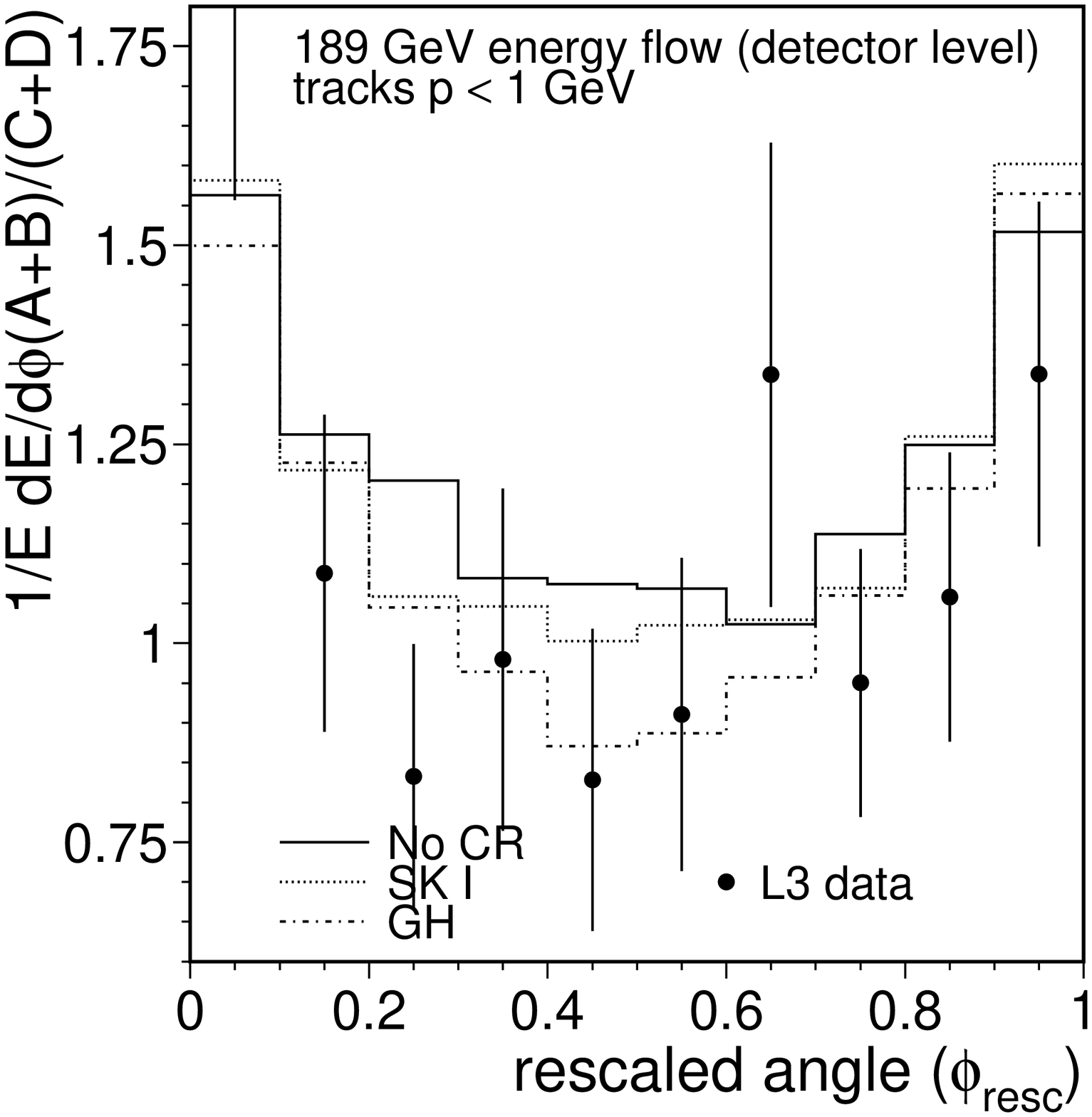,width=5.2cm,clip=}
\end{minipage}
\vspace{-0.5cm}
\caption{\small a) The particle flow (on the left) and b) the ratio of the
energy flow distribution (on the right)} 
\label{crc_l31}
\end{figure}

In the context of the string fragmentation model CRC should change the
particle production between different jets. This can be investigated via
particle and energy flow distributions. 
The particle flow describes the number of particles produced per angular unit
between 2 adjacent jets. In the energy flow histogram the entries are
weighted with the energy of the particles.
The analysis of the L3 collaboration [\cite{crc_l31}] uses a topological
event selection, afterwards the jets coming from one W boson are identified
and sorted. The angle between the jets is rescaled to be one.
Angles from 0 to 1 (A) and from 2 to 3 (B) in figure \ref{crc_l31}a
describe the region between the jets from the same W boson,
angles from 1 to 2 (C) and from 3 to 4 (D) describe the region between
jets from different W bosons.
The data show no significant deviation from the MC without CRC.
The ratio between the particle production of jets coming from the same and
from different W bosons 
$ \frac{particle flow (region A+B)}{particle flow (region C+D)}$
can be used as a more sensitive variable to CRC.
In figure \ref{crc_l31}b this ratio is shown for the energy flow
distribution of the low momentum particles ($p < 1$ $GeV$)
in comparison with one of the Sj\"ostrand/Khoze models (SK I)
as well as the somewhat extreme Gustafson/H\"akkinen model (GH)
[\cite{th_gh1}].
Although the errors are still quite large it seems that most of the data
points are systematically slightly lower than the standard MC prediction
without CRC. Nevertheless no evidence for CRC can be claimed within the
statistical significance. 

\section{Bose-Einstein Correlations (BEC)}
From quantum mechanics it is known that the wave function of a pair of
identical bosons must be symmetric. As a consequence the number of identical
bosons, produced close in phase space, is enhanced.
The so called Bose-Einstein correlations (BEC) are well established for
pions in $Z^o$ decays.
However it is not clear to what extent BEC are induced from the
space-time overlap of the decay products of the pair-produced W bosons.
From MC studies the shift in the W mass in the hadronic channel is estimated
to the order of $\Delta M_W \le \mathcal{O}$$(50~MeV)$.

\subsection{BEC Studies at ALEPH}
\begin{figure}[t]
\begin{minipage}{6.2cm}
\epsfig{file=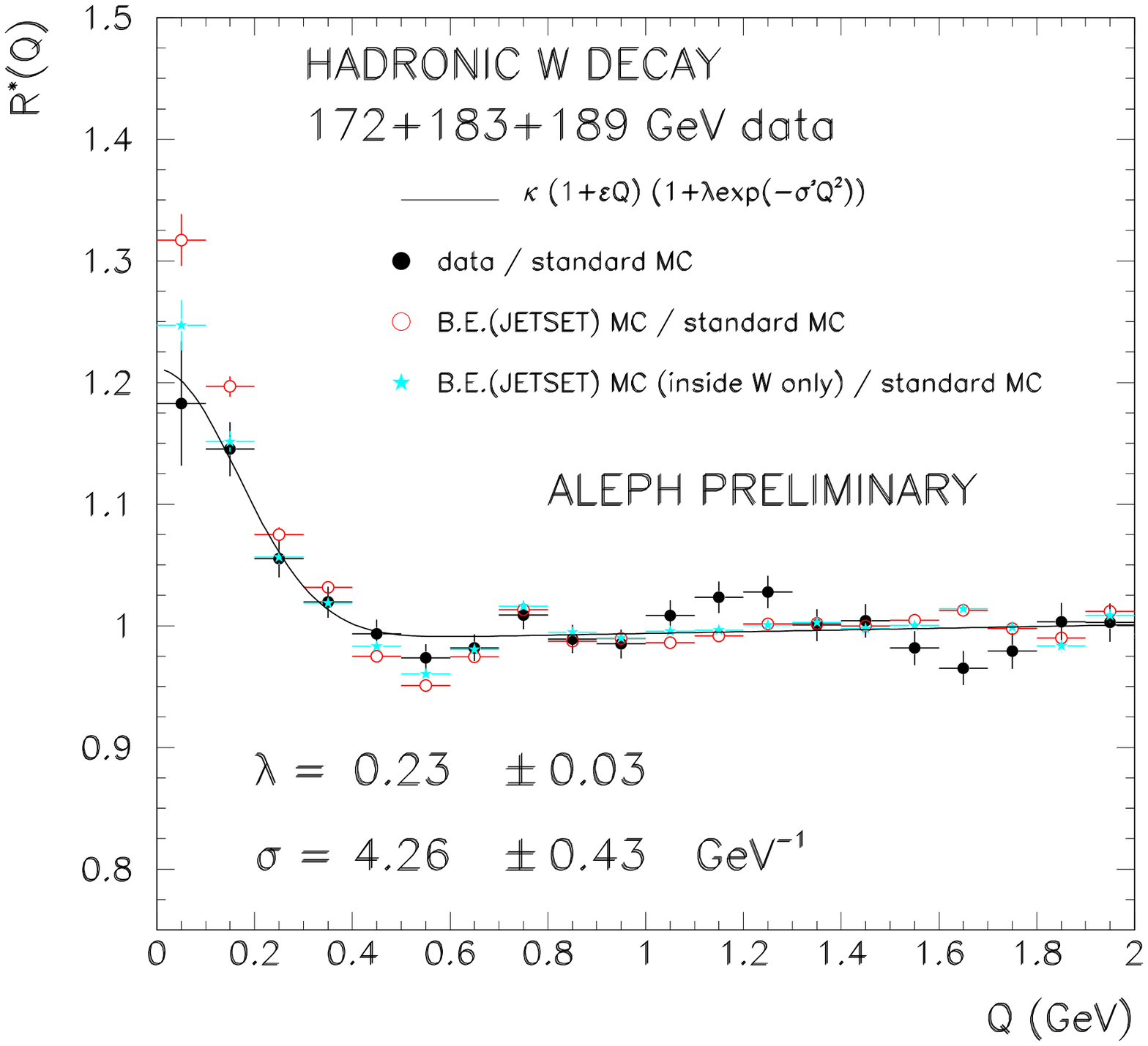,width=6.2cm,clip=}
\end{minipage}
\begin{minipage}{4.2cm}
\epsfig{file=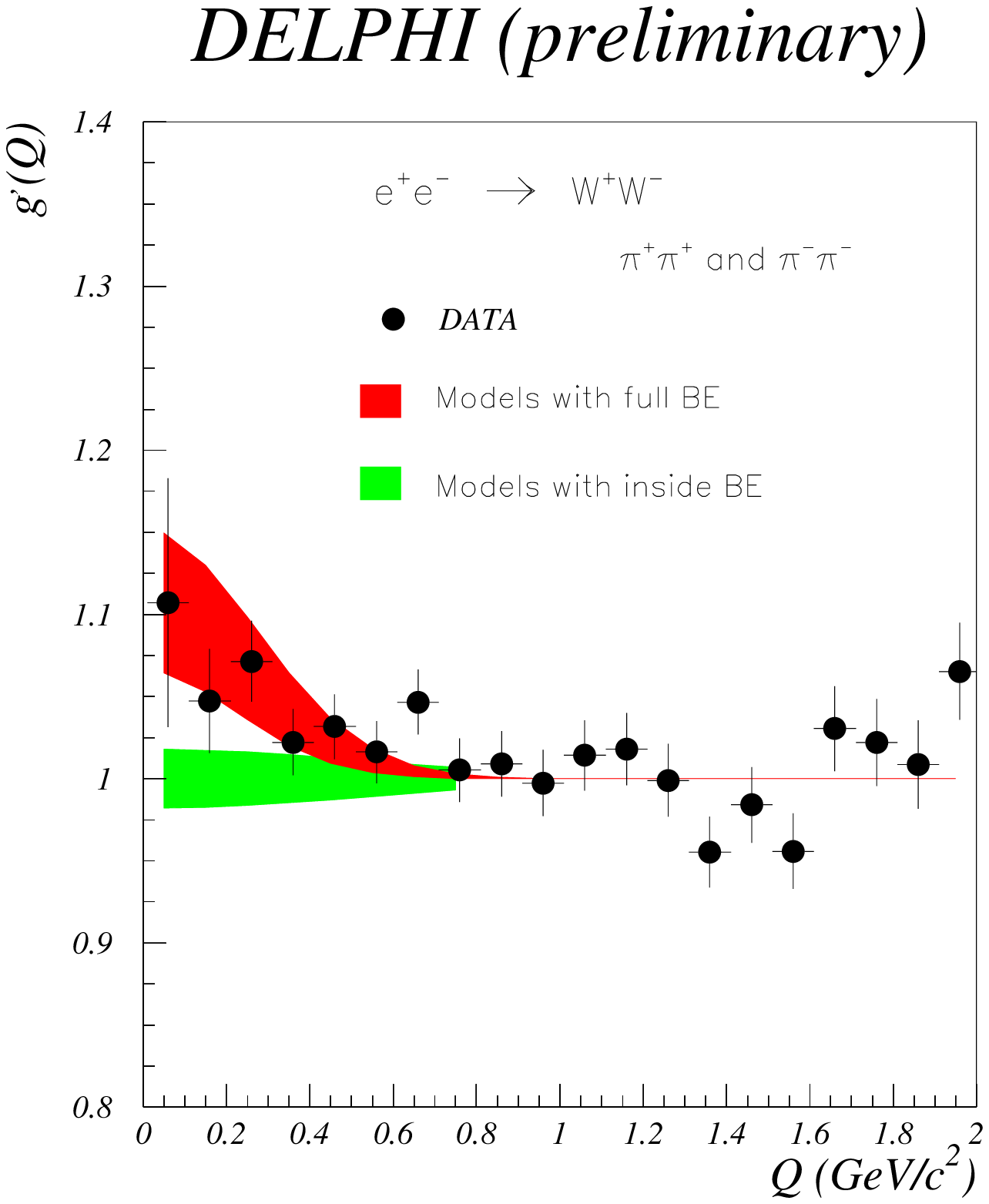,width=4.2cm,clip=}
\end{minipage}
\caption{\small a) The pion pair correlation function of ALEPH (left) and b)
of DELPHI (right) with MC predictions with and without BEC.}
\label{fig_bec_aleph1}
\end{figure}

The ALEPH collaboration uses the following double ratio [\cite{be_aleph1}]: \\
\centerline{
\begin{math}
R^*(Q) = \left( \frac{N_\pi^{++,--}(Q)}{N_\pi^{+-}(Q)} \right) ^{data} 
         {\Big /}
         \left( \frac{N_\pi^{++,--}(Q)}{N_\pi^{+-}(Q)} \right) ^{MC}_{noBEC}
\end{math}}
The ratio of the number of 'like-sign pion pairs' to 'unlike-sign pion pairs'
of the data is compared with a MC sample without BEC.
This ratio is parametrized with
$R^* (Q) = \kappa ( 1+\epsilon Q) ( 1 + \lambda e^{-\sigma ^2 Q^2})$.
$Q$ is the Lorentz-invariant momentum distance of
2 bosons, $\kappa$ defines the strength of the correlations and
$\sigma$ the source size of the boson emitter.
The double ratio and the fit to the data are shown in figure
\ref{fig_bec_aleph1}a. The data points are compared to MC with and without
BEC which is tuned and corrected at the $Z^o$ peak. BEC between
pions from different W bosons are disfavoured by the MC.

\subsection{BEC Studies at DELPHI}
The DELPHI collaboration uses an event mixing technique
[\cite{be_delphi1}] where they compare the number of 'like-sign pion pairs'
in the hadronic channel with the number of 'like-sign pion pairs' of a
hadronic event, which is built of the hadronic part of 2 semileptonic
events:\\
\centerline{
$g'(Q) = \left( \frac{N_{4q}(Q)}{N_{mix}(Q)} \right) ^{data} {\Big /}
         \left( \frac{N_{4q}(Q)}{N_{mix}(Q)} \right) ^{MC}_{noBEC}
$}\\
By definition there are no correlations between the pions of different W
bosons. The double ratio is parametrized as
$g'(Q) = 1 + \Delta e^{-k^2 Q^2}$.
The fit results in $\Delta = (7.3 \pm 2.5_{stat} \pm 1.8_{syst}) \%$.
If there are no BEC between pions from different W bosons the expectation
would be $\Delta = 0$. Therefore the BEC between the W bosons are preferred by
this analysis.

\subsection{BEC Studies at L3}
In the subtraction method of the L3 collaboration [\cite{be_l31}] a 2
particle density function is defined as
$\rho(p_1,p_2) = \frac{1}{N_{evt}} \frac{dn_{pairs}}{dQ}$.\\
The overall 2 particle density can be written as: \\
\centerline{
\begin{math}
\rho^{WW} = 2 \rho^W + 2 \rho^{WW}_{mix} \\
\rightarrow 
\hspace{0.5cm}
\Delta \rho = \rho^{WW} + 2 \rho^W + 2 \rho^{WW}_{mix}
\end{math}}\\
with $\Delta \rho = 0$ for no BEC between pion pairs from different W
bosons. 
The $\Delta \rho$ distributions for the like-sign and
unlike-sign charged pions are investigated as well as
$\delta \rho = \Delta \rho (\pm,\pm) - \Delta \rho (+,-)$ which should only
depend on BEC. No excess is observed in the distributions and BEC between
pions from different W bosons are disfavoured.

\subsection{BEC Studies at OPAL}
The OPAL collaboration uses the same double ratio as ALEPH but performs a
simultanous fit [\cite{be_opal1}] to the following event classes:\\
hadronic WW, semileptonic WW and $e^+e^- \rightarrow q \bar{q}$ events.
In table \ref{tab_bec_opal1} the fit result of the source size R and the
strength of the correlations $\lambda$ are shown.\\
With $\lambda = 0.05 \pm 0.67 \pm 0.35$ BEC between pions coming from
different W bosons cannot be established with the current level of
precision.
\begin{table}[t]
\begin{center}
\begin{tabular}{|c|c|c|}\hline
Parameter    & R(fm)                  & $\lambda$              \\ \hline \hline
same W       & $1.07\pm 0.07\pm 0.12$ & $0.69\pm 0.12\pm 0.06$ \\ \hline
diff W       & $1.51\pm 0.05\pm 0.35$ & $0.05\pm 0.67\pm 0.35$ \\ \hline
$(Z^o / \gamma)^*$ & $1.01\pm 0.08\pm 0.14$ & $0.43\pm 0.06\pm 0.08$ \\ \hline
\end{tabular}
\end{center}
\caption{\small The fit results of the OPAL collaboration}
\label{tab_bec_opal1}
\end{table}

\section{Summary}
Studies of final state interactions are an interesting field in itself and
may help to obtain a deeper understanding of the space-time development of
the fragmentation process.\\
The analyses on color reconnection show that all the investigated
distributions are compatible with standard MC predictions.
Nevertheless they are still limited by the statistics.
The particle and energy flow analysis of the L3 collaboration
looks promising as the statistics will be increased by at least a factor
of 3. With similar analyses of the other LEP collaborations the statistical
errors could be decreased by almost a factor of 4 with the final statistics.\\
The BEC analyses for the LEP collaborations use different methods and the
results are not yet conclusive.
More experimental effort is needed in this sector to understand BEC
resulting from the interference of the hadronic decay products of different W
bosons.\\
One should be aware that an interference between CRC and BEC may reduce or
enhance the overall effect.\\ 
The present experimental status of the W mass [\cite{gen_lep1}] is\\
\centerline{$M_W(q\bar{q}q\bar{q}) - M_W(q\bar{q} \ell \nu_{\ell}) = 
152 \pm 74~MeV/c^2$}\\
As the actual systematic error on final state interaction
is given with $58~MeV/c^2$ [\cite{gen_lep1}] it is important to understand
color reconnection and Bose-Einstein correlations because they are an
important systematic uncertainty on the W mass measurement in the hadronic
channel.

\section*{Acknowledgments}
I would like to thank all colleagues in the LEP collaborations who performed
the analyses and helped in preparing the talk.

\end{document}